\documentclass[journal,twoside]{IEEEtran}
\IEEEoverridecommandlockouts
\usepackage{cite}
\usepackage{amsmath,amssymb,amsfonts}
\usepackage{algorithmic}
\usepackage{graphicx}
\usepackage{textcomp}
\usepackage{xcolor}
\usepackage{subfigure}
\usepackage{flushend}
\usepackage{lipsum}
\usepackage{epstopdf}
\def\BibTeXfig1{{\rm B\kern-.05em{\sc i\kern-.025em b}\kern-.08em
    T\kern-.1667em\lower.7ex\hbox{E}\kern-.125emX}}
\begin{document}
\title{Hybrid Beamforming/Combining for Millimeter Wave MIMO: A Machine Learning Approach}
\author{Jiyun~Tao\IEEEauthorrefmark{1},
        Jing~Xing\IEEEauthorrefmark{1},
        Jienan~Chen\IEEEauthorrefmark{1}\IEEEmembership{Member,~IEEE},
        Chuan~Zhang\IEEEauthorrefmark{2}\IEEEmembership{Member,~IEEE},\\
        Shengli~Fu\IEEEauthorrefmark{3}\IEEEmembership{Senior Member,~IEEE}\\
        \IEEEauthorrefmark{1}University of Electronic Science and Technology of China, Chengdu, China (Email:Jesson.Chen@outlook.com)\\
        \IEEEauthorrefmark{2}National Mobile Communications Research Laboratory, Southeast University, Nanjing, China\\
        \IEEEauthorrefmark{3}Department of Electrical Engineering, University of North Texas, Denton, Texas, USA
       }

\markboth{Submitted to GlobalSIP}%
{Shell \MakeLowercase{\textit{et al.}}: Bare Demo of IEEEtran.cls for IEEE Journals}
\maketitle
\newcommand\blfootnote[1]{%
\begingroup
\renewcommand\thefootnote{}\footnote{#1}%
\addtocounter{footnote}{-1}%
\endgroup
}
\blfootnote{This work was supported by National Nature Science Foundation of China with Grant No. 61971107. Corresponding author is Jienan Chen (jesson.chen@outlook.com)}

\begin{abstract}
Hybrid beamforming (HB) has emerged as a promising technology to support ultra high transmission capacity and with low complexity for Millimeter Wave (mmWave) multiple-input and multiple-output (MIMO) system. However, the design of digital and analog beamformer is a challenge task with non-convex optimization, especially for the multi-user scenario. Recently, the blooming of deep learning research provides a new vision for the signal processing of communication system. In this work, we propose a deep neural network based HB for the multi-User mmWave massive MIMO system, referred as DNHB. The HB system is formulated as an autoencoder neural network, which is trained in a style of end-to-end self-supervised learning. With the strong representation capability of deep neural network, the proposed DNHB exhibits superior performance than the traditional linear processing methods. According to the simulation results, DNHB outperforms about 2 dB in terms of bit error rate (BER) performance compared with existing methods.

\end{abstract}


\section{Introduction}
The massive multiple-input multiple-output (MIMO) with Millimeter Wave (mmWave) can provide a high spatial gain and diversity gain for the high data transmission, which is considered as a key technology for the future wireless communication system \cite{wenxian1}. The HB technology, which consists of digital and analog beamformer, is with low hardware and computation complexity and maintains high data transmission capacity. \cite{wenxian2}

However, the global optimization of digital and analog beamformer is still a challenge task \cite{wenxian9}. The analog beamformer/combiner is implemented with a constant modulus constraint phase shifter network \cite{wenxian10}. Besides that, the analog beamformer/combiner and digital beamformer/combiner are coupled. Consequently, the joint optimization of HB system is an non-convex problem.  This problem is even more difficult for the multi-user scenario, where the beamformer/combiner design for each user can only be optimized separatively \cite{wenxian11}.

The existing methods \cite{wenxian12, wenxian13, wenxian14,wenxian15} to solve the challenges and achieve feasible near-optimal solutions attempt to approximate the full digital optimal beamformer by decoupled the design of analog and digital stage. It first fixes one processing stage, i.e., analog beamformer, and optimizes the digital stage with the optimization goal of full digital beamforming matrix. Then the digital beamformer is fixed and the analog stage is optimized. The two processes are performed iteratively until the algorithm is converged. However, the solution is sub-optimal by the limitation of analog beamformer based on matrix decomposition methods. To this end, the result will prone to be a local-optimum point for the separating optimizing digital and analog part.

As alternative optimization tools, machine learning and artificial intelligence (AI) provide new approaches for solving
the over-complicated problems in communication systems, which have been applied in intelligent radio network \cite{wenxian3,wenxian4,wenxian5}, backhaul optimization for mmWave system \cite{wenxian6} and signal processing for physical layer,  \cite{wenxian7, wenxian8} and network traffic analysis \cite{R1,R2} etc. As the communication problem just gets what you transmit, it is exactly as the same as the expectation of autoencoder neural network which hopes outputs equal inputs. Hence, to address the above critical challenges, we introduce a deep neural network based HB design method by mapping HB multi-user system as an autoencoder (AE) neural network, referred as DNHB.

The contributions of this work are summarized as follows .
\begin{itemize}
  \item  Deep autoencoder neural network based self-supervised learning. Instead of applying the data and labels from matrix decomposition method for training, we integrate the CSI matrix into a hidden layer of the considered AE neural network. By training the DNN in an end-to-end style, the proposed AE neural network HB can break through the performance limit of the existing matrix decomposition method.
  \item Supporting multi-user neural network scenario. Compared with our previous work, we propose a splitting neural network to support multi-user massive MIMO system scenario. After the channel network layer, we decompose the network to several sub-layer networks to match the practical multi-user MIMO system.
  \item Superior performance over the existing methods. The proposed design outperforms the existing methods about 2 dB in bit error rate (BER) performance.
\end{itemize}

The organization of this paper is as follows. In section II, the multi-user mmWave massive MIMO HB system module is illustrated and the optimization problem is clarified. In section III, multi-user mmWave massive MIMO the AE based HB system structure is proposed. The map from traditional structure to the neural network is detailed explained. The following sections present the comparison of simulation results in BER with other methods. The proposed method can get 2-3 dB gain in the contrast of traditional algorithms.

The notation in this paper is introduced as below $\mathbf{A}$ represents the matrix and $\mathbf{A}(i,j)$ represents the $(i,j)-$th entry of a matrix $\mathbf{A}$.  $\left| \cdot  \right|$ is the modulus of a complex number and ${{\left\| \cdot  \right\|}_{F}}$ is the Frobenius norm. For a vector or matrix, the superscripts ${{(\cdot )}^{T}}$, ${{(\cdot )}^{*}}$ and ${{(\cdot )}^{H}}$ represent transpose, complex conjugate and complex conjugate transpose, respectively. $\Re (\cdot )$ is the real-part operator and $\Im (\cdot )$ is the imaginary-part operator. $E[\cdot]$ denotes expectation, and ${{\mathbf{I}}_{K}}$ is the $K\times K$ identity matrix.
\section{System model and problem formulation}
\subsection{system model}

\begin{figure}[t]
\centerline{\includegraphics[width=3.5in]{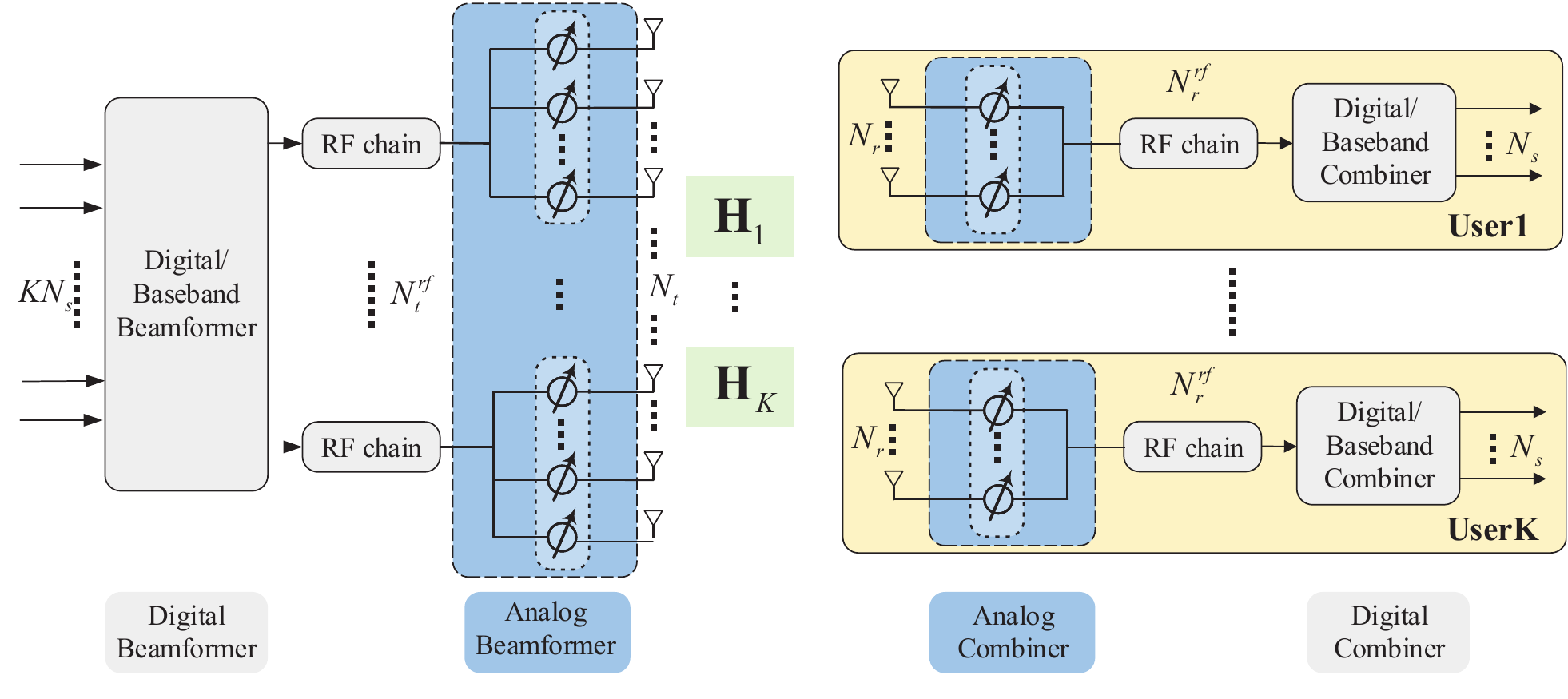}}
\caption{Illustration of a narrowband downlink single-cell multi-user mmWave massive MIMO HB system.}
\label{fig1}
\end{figure}

A typical narrowband downlink single-cell multi-user mmWave massive MIMO system is as shown in Fig.~\ref{fig1}. A base station (BS) is equipped with ${{N}_{t}}$ transmit antennas, and $N_{t}^{rf}$ radio frequency (RF) chains. Each RF chains serves $K$ users, which is equipped with ${{N}_{r}}$ receiver antennas and $N_{r}^{rf}$ RF chains [1]. The number of independent data streams is ${{N}_{s}}$, which means that total $K{{N}_{s}}$ data streams are processed by the BS. To guarantee the quality with the limited number of RF chains, the number of the transmitted streams is constrained by $K{{N}_{s}}\le N_{t}^{rf}\le {{N}_{t}}$ for the BS and ${{N}_{s}}\le N_{r}^{rf}\le {{N}_{r}}$ for each user.

At the BS, the transmitted symbols are assumed to be processed by a $N_{t}^{rf}\times K{{N}_{s}}$ digital beamformer $ {{\mathbf{F}}_{\text{D}}}$ and then by an analog beamformer ${{\mathbf{F}}_{\text{A}}}$ of dimension ${{N}_{t}}\times N_{t}^{rf}$ to construct the final transmitted signal. Notably, the digital beamformer ${{\mathbf{F}}_{\text{D}}}$ enables both amplitude and phase modification, while only phase changes (phase-only control) can be realized by analog beamformer  ${{\mathbf{F}}_{\text{A}}}$ with only phase shifters. To this end, each element of ${{\mathbf{F}}_{\text{A}}}$ is normalized to satisfy ${{\left| {{\mathbf{F}}_{\text{A}}}(i,j) \right|}^{2}}=1$. Mathematically, the transmitted signal can be written as
\begin{equation}
{\bf{x}} = {\bf{Fs}} = {{\bf{F}}_{\rm{A}}}{{\bf{F}}_{\rm{D}}}{\bf{s}} = \sum\limits_{k = 1}^K {{{\bf{F}}_{\rm{A}}}{{\bf{F}}_{{\rm{D}}k}}{{\bf{s}}_k}},
\label{equ1}
\end{equation}
where $\mathbf{F}={{\mathbf{F}}_{\text{A}}}{{\mathbf{F}}_{\text{D}}}$ denotes the HB matrix with ${{\bf{F}}_{\rm{D}}}{\rm{ = }}[ {{{\bf{F}}_{{\rm{D1}}}}, {{\bf{F}}_{{\rm{D2}}}},...,{{\bf{F}}_{{\rm{D}}K}}} ]$. ${{\mathbf{F}}_{\text{D}k}}$ is a digital beamformer matrix for $k=1,2,...,K$, and $\mathbf{s}\in {{\mathbb{C}}^{K{{N}_{s}}\times 1}}$ is the vector of signal symbols which is the concatenation of each user's data stream vector such as $\mathbf{s}={{[ \mathbf{s}_{1}^{T},\mathbf{s}_{2}^{T},...,\mathbf{s}_{K}^{T} ]}^{T}}$, where ${{\mathbf{s}}_{k}}$ denotes the data stream vector for user $k$. It is assumed that $E [ \mathbf{s}{{\mathbf{s}}^{H}} ]={{\mathbf{I}}_{K{{N}_{s}}}}$. For user \emph{k}, the received signal can be modeled as
\begin{equation}
{{\bf{y}}_k} = {{\bf{H}}_k}{{\bf{F}}_{\rm{A}}}{{\bf{F}}_{{\rm{D}}k}}{{\bf{s}}_k} + {{\bf{H}}_k}\sum\limits_{l \ne k} {{{\bf{F}}_{\rm{A}}}{{\bf{F}}_{{\rm{D}}l}}{{\bf{s}}_l}}  + {{\bf{n}}_k},
\label{equ2}
\end{equation}
where ${{\mathbf{H}}_{k}}\in {{\mathbb{C}}^{{{N}_{r}}\times {{N}_{t}}}}$ is the channel matrix for the $k$-th user and ${{\mathbf{n}}_{k}}\in {{\mathbb{C}}^{{{N}_{s}}\times 1}}$ is the vector of i.i.d. $\mathcal{CN}(0,{{\sigma }^{2}})$ additive complex Gaussian noise. The received signal after beamformer at $k$-th user is given by
\begin{equation}
\begin{split}
{{\bf{\tilde s}}_k}=& \underbrace {{\bf{W}}_{{\rm{D}}k}^H{\bf{W}}_{{\rm{A}}k}^H{{\bf{H}}_k}{{\bf{F}}_{\rm{A}}}{{\bf{F}}_{{\rm{D}}k}}{{\bf{s}}_k}}_{{\rm{desired \ signals}}} + \underbrace {{\bf{W}}_{{\rm{D}}k}^H{\bf{W}}_{{\rm{A}}k}^H{{\bf{H}}_k}\sum\limits_{l \ne k} {{{\bf{F}}_{\rm{A}}}{{\bf{F}}_{{\rm{D}}l}}{{\bf{s}}_l}} }_{{\rm{effective \ interference }}} \\
&+ \underbrace {{\bf{W}}_{{\rm{D}}k}^H{\bf{W}}_{{\rm{A}}k}^H{{\bf{n}}_k}}_{{\rm{effective \ noise}}},
\label{equ3}
\end{split}
\end{equation}
where ${{\mathbf{W}}_{\text{A}k}}\in {{\mathbb{C}}^{{{N}_{r}}\times N_{r}^{rf}}}$ is the analog combining matrix and ${{\mathbf{W}}_{\text{D}k}}\in {{\mathbb{C}}^{N_{r}^{rf}\times {{N}_{s}}}}$ is the digital combining matrix for the $k$-th user. Since ${{\mathbf{W}}_{\text{A}k}}$ is also implemented by the analog phase shifters, all elements of ${{\mathbf{W}}_{\text{A}k}}$ should have the constant amplitude such that ${{\left| {{\mathbf{W}}_{\text{A}k}}(i,j) \right|}^{2}}=1$. In this paper, it is assumed that perfect channel state information (CSI) is available at both the transmitter and receiver and that there is perfect synchronization between them.
\subsection{Problem formulation}
In this work, we employ the sum-MSE \cite{wordwx2}, \cite{wordwx3} of all users and all streams as the performance measure and optimization objective for the joint transmit and receive HB design, which is defined as
\begin{equation}
{\rm{MSE}} \buildrel \Delta \over = \sum\limits_{k = 1}^K {{\rm{MS}}{{\rm{E}}_k}}  = \sum\limits_{k = 1}^K {{E}\left\{ {{{\left\| {{{\bf{s}}_k} - {{{\bf{\tilde s}}}_k}} \right\|}^2_F}} \right\}} ,
\label{equ4}\\
\end{equation}
where $ {{\rm{MS}}{{\rm{E}}_k}}$ denotes the MSE of the $k$-th user. By substituting (\ref{equ3}) into (\ref{equ4}) and irrespective of effective interference, we have
\begin{equation}
\begin{split}
{\rm{MSE}} \buildrel \Delta \over = &\sum\limits_{k = 1}^KE\{ ||{{\bf{s}}_k} - ({\bf{W}}_{{\rm{D}}k}^H{\bf{W}}_{{\rm{A}}k}^H{\bf{H}}{{\bf{F}}_{\rm{A}}}{{\bf{F}}_{{\rm{D}}k}}{{\bf{s}}_k}  \\&+ {\bf{W}}_{{\rm{D}}k}^H{\bf{W}}_{{\rm{A}}k}^H{{\bf{n}}_k})|{|^2_F}\} .
\end{split}
\label{equ5}
\end{equation}\\
Finally, the optimization problem in the narrowband scenario for multi-user can be formulated as
\begin{equation}
\begin{split}
\mathop {{\mathop{\rm minmize}\nolimits} }\limits_{{{\bf{F}}_{{\rm{D}}k}},{{\bf{F}}_{\rm{A}}},{\bf{W}}_{{\rm{A}}k}^H,{\bf{W}}_{{\rm{D}}k}^H} \sum\limits_{k = 1}^K {{\rm{MS}}{{\rm{E}}_k}} \\
{\rm{subject \; to \;  tr}}\left\{ {{{\bf{F}}_{\rm{A}}}{{\bf{F}}_{\rm{D}}}{\bf{F}}_{\rm{D}}^H{\bf{F}}_{\rm{A}}^H} \right\} &\le P,\\
{\;}{\left| {{{\bf{F}}_{\rm{A}}}(i,j)} \right|^2} = 1,&\forall i,j,\\
{\;}{\left| {{{\bf{W}}_{{\rm{A}}k}}(i,j)} \right|^2} = 1,&\forall i,j,k,
\end{split}
\label{equ6}
\end{equation}
where $P$ is the total power budget at the BS. Obviously, the optimization problem is non-convex and it is intractable to obtain global optima for similar constrained joint optimization problems \cite{wordwx4}.
\section{AE based HB design}

\begin{figure}[t]
\centerline{\includegraphics[width=3.3in]{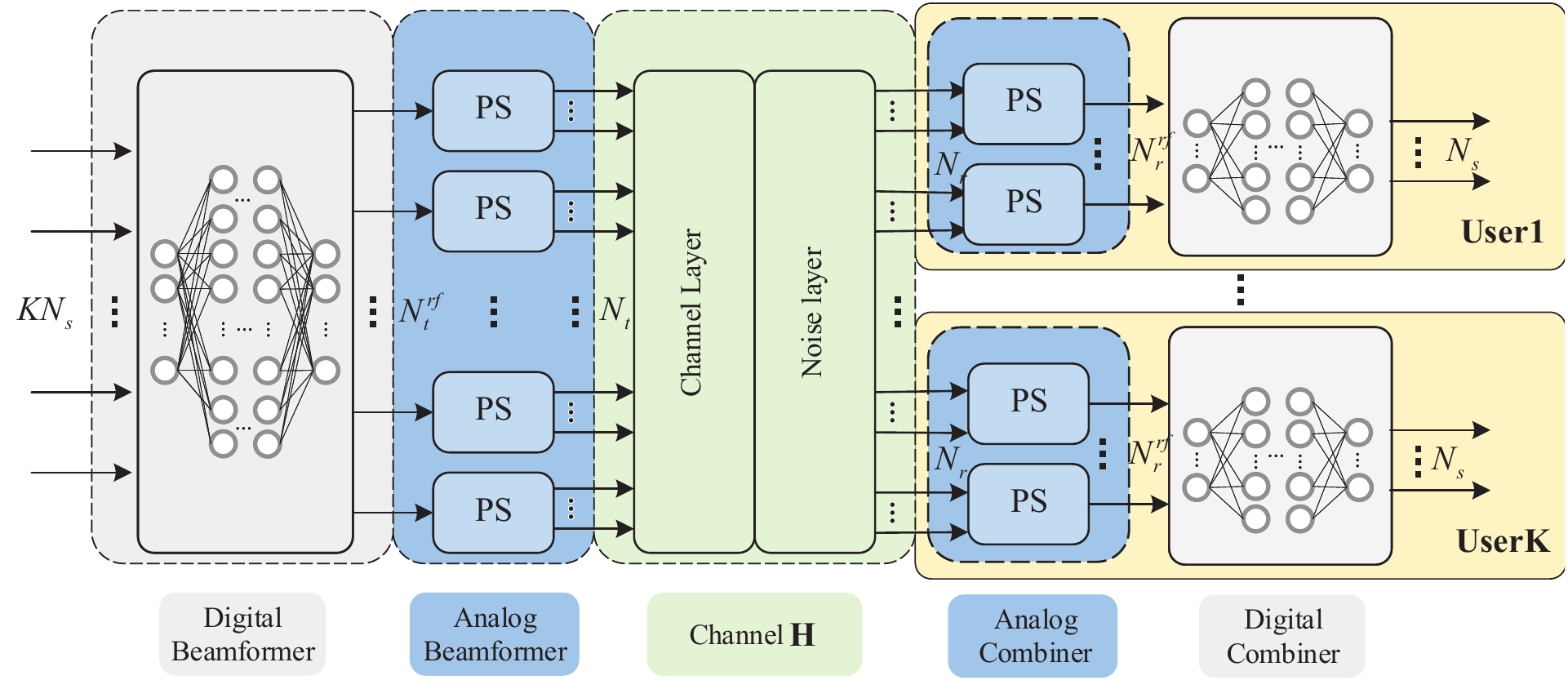}}
\caption{A simplified hardware block diagram of downlink single-cell multi-user mmWave massive MIMO autoencoder based HB system.}
\label{fig2}
\end{figure}

AE neural network is a subclass of the artificial neural network used for unsupervised learning. AE neural network was firstly proposed in data compression, by learning a presentation for a set of data, such as images. Traditional AE neural networks are designed to recovery the original data from compressed information, which is similar to the optimization problem as writing (\ref{equ6}). In this work, we propose a neural HB network by mapping the hardware block diagram of downlink single-cell multi-user mmWave massive MIMO system shown in Fig.~\ref{fig2} .

\subsection{Digital beamformer/combiner design}
For digital beamformer/combiner, the input complex signal $\mathbf{x}$ of the neural network is divided into two parts, the real part $\Re (\mathbf{x})$ and the imaginary part $\Im (\mathbf{x})$. Therefore, the input signal of the fully connected neural network can be written as $\mathbf{X}=\left[ \Re (\mathbf{x}),\Im (\mathbf{x}) \right]$. The function for a one layer fully connected neural network is given by
\begin{equation}
\begin{split}
{\bf{Y}} =& [\Re ({\bf{y}}),\Im ({\bf{y}})] = [\sigma ({{\bf{W}}_\Re }\Re ({\bf{x}})  - {{\bf{W}}_\Im }\Im ({\bf{x}}) +{{\bf{b}}_\Re }),\\&
\sigma ({{\bf{W}}_\Re }\Im ({\bf{x}})  + {{\bf{W}}_\Im }\Re ({\bf{x}}) + {{\bf{b}}_\Im }))],
\end{split}
\label{equ7}
\end{equation}
where $\bf{Y}$ is the concatenation of the real part $\Re (\mathbf{y})$, and the imaginary part $\Im (\mathbf{y})$ of the processed complex signal $\bf{y}$. The weights of the real and imaginary channel can be stated as ${{\mathbf{W}}_{\Re }}$ and ${{\mathbf{W}}_{\Im }}$, respectively. The biases of the real and imaginary channel can be stated as ${{\mathbf{b}}_{\Re }}$ and ${{\mathbf{b}}_{\Im }}$, respectively. Here, the symbol $\sigma (\cdot )$ denotes the activation function.

Digital beamformer/combiner can adjust both the phase and the amplitude of the original signals without limitations. We employ two \emph{n}-layers fully connected neural networks to implement baseband beamformer/combiner, respectively. Concisely, the processed signal of the digital beamformer is represented as
\begin{equation}
{{\bf{S}}_{\rm{D}}} = {\bf{f}}_t^n({\bf{S}};{\alpha _t}),
\label{equ8}
 \end{equation}
where ${{\mathbf{S}}_{\text{D}}}=\left[ \Re ({{\mathbf{s}}_{\text{D}}}),\Im ({{\mathbf{s}}_{\text{D}}}) \right]$. The output complex signal ${{\mathbf{s}}_{\text{D}}}$ of the digital beamformer neural network can be obtained by combining ${{\mathbf{S}}_{\text{D}}}$. Here, $\mathbf{f}_{t}^{n}$ represents the concatenation of \emph{n}-layers fully connected neural network and ${{\alpha }_{t}}$ represents the parameters set of the real channel and imaginary channel in the \emph{n}-layer digital beamformer neural network.

Similarly, the receivers signal after baseband combining can be stated as
\begin{equation}
{\bf{\tilde S}} = {\bf{f}}_r^n({{\bf{S}}_{\rm{A}}};{\alpha _r}),
\label{equ9}
 \end{equation}
where the output of the digital combiner neural network can be written as $\mathbf{\tilde{S}}=\left[ \Re (\mathbf{\tilde{s}}),\Im (\mathbf{\tilde{s}}) \right]$. Here, $\mathbf{\tilde{s}}={{\left[ \mathbf{\tilde{s}}_{1}^{T},\mathbf{\tilde{s}}_{2}^{T},...,\mathbf{\tilde{s}}_{K}^{T} \right]}^{T}}$ denotes the concatenation of each user's signal symbols processed by HB system. For each element ${{\mathbf{\tilde{s}}}_{k}}$, it denotes the received data stream vector for user $k$. Here, ${{\mathbf{S}}_{\text{A}}}\text{=}\left[ \Re ({{\mathbf{s}}_{\text{A}}}),\Im ({{\mathbf{s}}_{\text{A}}}) \right]$ represents the output of the analog combiner neural network. The complex signal processed by analog combiner be written as ${{\mathbf{s}}_{\text{A}}}$. Besides, ${{\alpha }_{r}}$ represents the parameters set of the real channel and imaginary channel in the $n$-layer digital combiner neural network.
\subsection{Analog beamformer/combiner design}
The analog beamformer/combiner neural network should also satisfy the constraints of analog phase
shifters to match the practical hardware scheme. For fully connected beamformer, each RF chain is connected to all ${{N}_{r}}$ antennas via phase shifter neural network. The transmit complex signals processed by analog beamformer can be written as
\begin{equation}
\begin{array}{l}
{{\bf{s}}^t} = \rho {[s_1^t,s_2^t,...,s_{{N_t}}^t]^T}\\
 = \rho {[\sum\limits_{p = 1}^{N_t^{rf}} {{s_{{\rm{D,1}}}}{e^{j\theta _{p,1}^t}},} \sum\limits_{p = 1}^{N_t^{rf}} {{s_{{\rm{D,2}}}}{e^{j\theta _{p,2}^t}},} ...,\sum\limits_{p = 1}^{N_t^{rf}} {{s_{{\rm{D,}}N_t^{rf}}}{e^{j\theta _{p,{N_t}}^t}}} ]^T},
\end{array}
\label{equ10}
 \end{equation}
where $s_{n}^{t}$, $n\in \{1,2,...,{{N}_{t}}\}$ denotes the transmitted signal of the $n$-th antenna, and ${{s}_{\text{D,}p}}$, $p\in \{1,2,...,N_{t}^{rf}\}$ denotes the $p$-th RF chain signal processed by digital beamformer. To meet the transmit
power constraint, the power control parameter $\rho $ is set as
\begin{equation}
\rho  = {(P\sum\limits_{n = 1}^{{N_t}} {{{\left| {s_n^t} \right|}^2}} )^{ - 1/2}}.
\label{equ11}
 \end{equation}

For user $k$, the processed signal by analog combiner can be modeled as

\begin{equation}
\begin{split}
{{\bf{s}}_{{\rm{A,}}k}}=& {[{s_{{\rm{A,}}k,1}},{s_{{\rm{A,}}k,2}},...,{s_{{\rm{A,}}k,N_r^{rf}}}]^T}\\
 =& [\sum\limits_{m = 1}^{{N_r}} {s_{k,1}^r{e^{j\theta _{k,m,1}^r}}} ,\sum\limits_{m = 1}^{{N_r}} {s_{k,1}^r{e^{j\theta _{k,m,2}^r}}} ,...,\\&\sum\limits_{m = 1}^{{N_r}} {s_{k,1}^r{e^{j\theta _{k,m,N_r^{rf}}^r}}} ]^T,
\end{split}
\label{equ12}
 \end{equation}
where ${{s}_{\text{A,}k,q}}$, $q\in \{1,2,...,N_{r}^{rf}\}$ denotes the $q$-th RF chain's signal of the user $k$. Here, $\theta _{k,m,q}^{r}$, $m\in \{1,2,...,{{N}_{r}}\}$ denotes the phase parameter between the $m$-th receive antenna and the $q$-th RF chain for the user $k$.

Notice that the formulas (\ref{equ10}) and (\ref{equ12}) involve multiplication of complex number such as $x{{e}^{j\theta }}$, combining with Euler's formula $x{{e}^{j\theta }}$ is equivalent to
\begin{equation}
y = x{e^{j\theta }} = x(cos\theta  + j\sin \theta ).
\label{equ13}
 \end{equation}

In phase shifter neural network, the multiplication in (\ref{equ13}) can be stated as
\begin{equation}
\begin{split}
Y &=[\Re (y),\Im (y)]\\&
 = [(\Re (x)cos\theta  - \Im (x)\sin \theta ),(\Re (x)\sin \theta  + \Im (x)cos\theta )].
\end{split}
\label{equ14}
 \end{equation}

 Similar to the writing (\ref{equ8}) and (\ref{equ9}) the output of the analog beamformer/combiner can be represented respectively as
 \begin{equation}
\begin{array}{l}
{{\bf{S}}^t} = {g_t}({{\bf{S}}_{\rm{D}}};{{\bf{\theta }}^t}),\\
{{\bf{S}}_{\rm{A}}} = {g_r}({{\bf{S}}^r};{{\bf{\theta }}^r}),
\end{array}
\label{equ15}
 \end{equation}
 where ${{g}_{t}}(\cdot )$/${{g}_{r}}(\cdot )$ denotes the analog beamformer/combiner neural network. The phase parameters are represented as ${{\mathbf{\theta }}_{t}}/{{\mathbf{\theta }}_{r}}$.

 \subsection{Channel transmission}

The channel transmission and noise adding process are realized by neural network with fixed parameters, for the any $k$-th user, which is stated as
\begin{equation}
\begin{split}
\mathbf{S}_{k}^{r}&=[\Re (\mathbf{s}_{k}^{r}),\Im (\mathbf{s}_{k}^{r})] \\
 & =[\Re ({{\mathbf{s}}^{t}}{{\mathbf{H}}_{k}}+{{\mathbf{n}}_{k}}),\Im ({{\mathbf{s}}^{t}}{{\mathbf{H}}_{k}}+{{\mathbf{n}}_{k}})] \\
\end{split}
\label{equ16}
 \end{equation}
 where ${{\mathbf{s}}^{r}}={{\left[ {{\left( \mathbf{s}_{1}^{r} \right)}^{T}},{{\left( \mathbf{s}_{2}^{r} \right)}^{T}},\ldots ,{{\left( \mathbf{s}_{K}^{r} \right)}^{T}} \right]}^{T}}$. ${{\mathbf{s}}^{t}}$ is the transmitted signal at BS, and $\mathbf{s}_{k}^{r}$ is the received signal at user $k$.  The CSI matrix between the user $k$ with the BS is represented as ${{\mathbf{H}}_{k}}$. ${{\mathbf{n}}_{k}}$ represents the corresponding noise vector of i.i.d. $\mathcal{C}\mathcal{N}\left( 0,\sigma _{k}^{2} \right)$.

 \subsection{Optimization Problem Formation}
  In the multi-user system, the sum-MSE (\ref{equ4}) is regarded as the reconstruction error. Thus, the loss function of DNHB is
\begin{equation}
\begin{split}
{\cal L} =& {\cal L}\left( {{\bf{S}},{\bf{\tilde S}}} \right) = \sum\limits_{k = 1}^K {{E}{{\left\| {{{\bf{s}}_k} - {{{\bf{\tilde s}}}_k}} \right\|}^2_F}}  \\=& \sum\limits_{k = 1}^K {{E}\left[ {{{\left\| {\Re \left( {{{\bf{s}}_k}} \right) - \Re \left( {{{{\bf{\tilde s}}}_k}} \right)} \right\|}^2_F} + {{\left\| {\Im \left( {{{\bf{s}}_k}} \right) - \Im \left( {{{{\bf{\tilde s}}}_k}} \right)} \right\|}^2_F}} \right]}
\end{split}
\label{equ17}
 \end{equation}

\section{Simulation}
In the simulation experiments, the mmWave propagation channel is based on a geometry channel model \cite{wordwx1}. The configuration of the MIMO system simulation is set as, ${{N}_{t}}=64$ at BS, ${{N}_{r}}=16$, $N_{r}^{rf}=2$ and $N_s=2$ at each user.  The number of channel clusters is 2 and 2 rays with each cluster.

\begin{figure}[t]
\centerline{\includegraphics[width=2.7in]{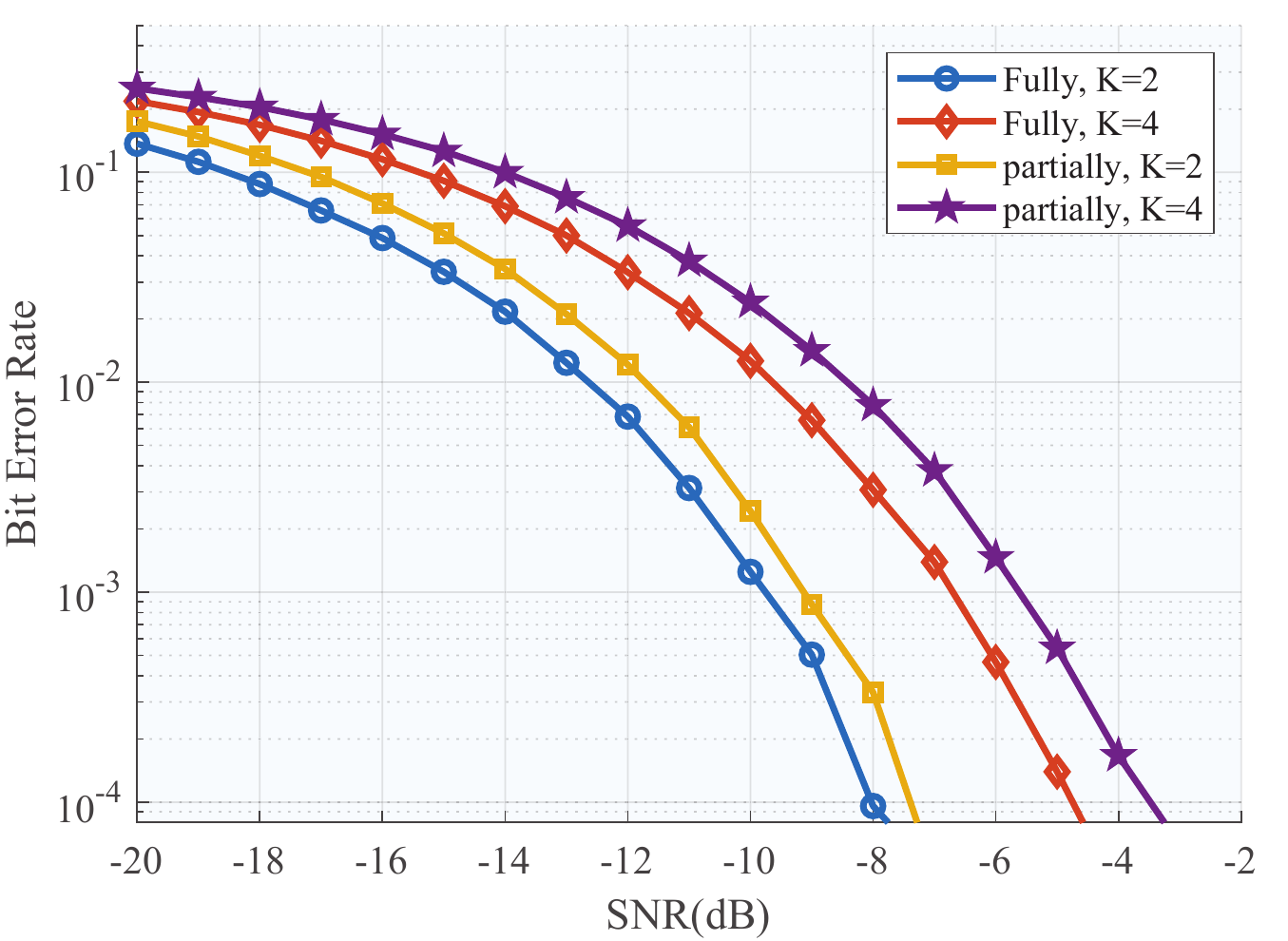}}
\caption{Illustration of the BER comparison with different number of users $K=2,4$ DNHF. The parameters are set as $N_r^{rf}=2$, $N_t^{rf} = KN_r^{rf}$  and the total data streams $KN_s=2\times K$}
\label{fig4}
\end{figure}

Fig.~\ref{fig4} shows the BER comparison of the different number of users $K=2,4$ in fully connected and partially connected analog beamformer/combiner. As it depicts, the fully connected analog beamformer/combiner has 3dB better performance than partially connected analog beamformer/combiner in BER.  The partially connected analog beamformer/combiner has more constriction due to its structure. And the 2 users has superior performance in comparison than 4 users, as the information between different users cannot be transmitted and exchanged in the multi-user system. With the increases in number of users is in the system, its performance is suggested to get worse.

\begin{figure}
\centerline{\includegraphics[width=2.7in]{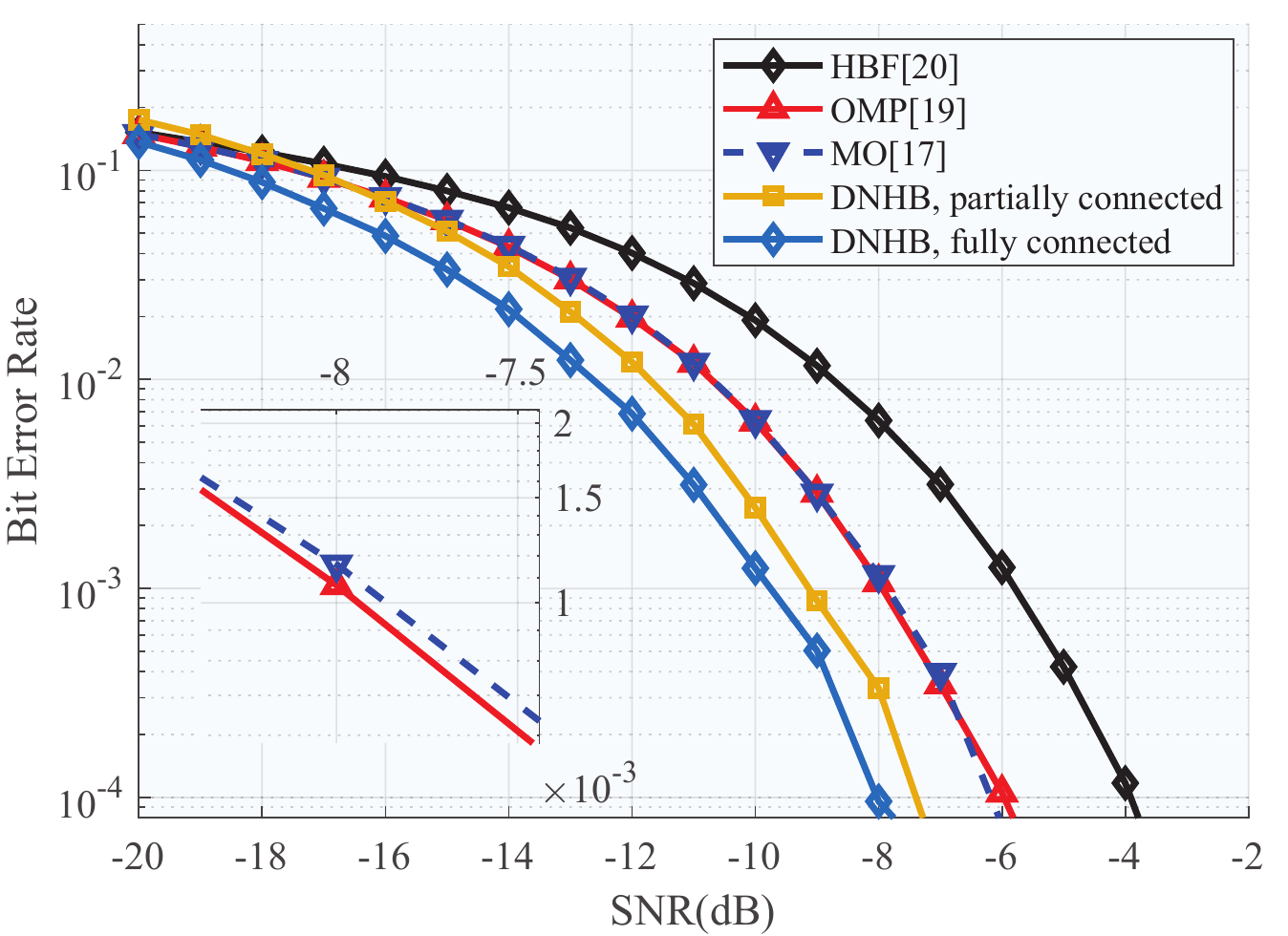}}
\caption{Illustration of the comparison with some existing methods. The parameters are set as $K=2$, $N_t^{rf} = N_r^{rf}=2$ and the totally data streams $KN_s=4$.}
\label{fig5}
\end{figure}

The Fig.~\ref{fig5} is the comparison with other exsiting beamforming methods, such as the conventional orthogonal matching pursuit based algorithm in \cite{OMP} (labelled with 'OMP') and one conventional HB algorithm in \cite{HBF} (labelled with 'HBF') and the manifold optimization HB algorithm in \cite{wordwx2} (labelled with 'MO'). It illustrates the BER performance comparison when $K=2$. The proposed DNHB algorithm has $2\sim3$dB performance  advantages.

\section{Conclusion}
In this paper, a new vision on the HB design is proposed in the perspective of autoencoder neural network. Compared with traditional matrix decomposition, the AE neural network provides a strong representation ability to map the non-convex HB design to a network training process. The method exhibits significantly superior performance than the traditional linear matrix decomposition methods, in terms of BER. The proposed work provides a new vision on the HB design, which may have great potential in guiding designs in the future intelligent radio.
\clearpage
\bibliographystyle{unsrt}
\bibliography{ref}

\begin{thebibliography}{10}

\bibitem{wenxian1}
Yikun Yu, Peter G.~M. Baltus, and Arthur H.~M. van Roermund.
\newblock Millimeter-{Wave} {Wireless} {Communication}.
\newblock In Yikun Yu, Peter~G.M. Baltus, and Arthur~H.M. van Roermund,
  editors, {\em Integrated 60GHz {RF} {Beamforming} in {CMOS}}, Analog
  {Circuits} and {Signal} {Processing}, pages 7--18. Springer Netherlands,
  Dordrecht, 2011.

\bibitem{wenxian2}
S~Mumtaz, J~Rodriguez, and Linglong Dai.
\newblock {\em {mmWave} {Massive} {MIMO}: {A} {Paradigm} for 5G}.
\newblock 2016.

\bibitem{wenxian9}
D.P. Palomar, J.M. Cioffi, and M.A. Lagunas.
\newblock Joint tx-rx beamforming design for multicarrier mimo channels: a
  unified framework for convex optimization.
\newblock {\em IEEE Transactions on Signal Processing}, 51(9):2381--2401,
  September 2003.

\bibitem{wenxian10}
Andreas~F. Molisch, Vishnu~V. Ratnam, Shengqian Han, Zheda Li, Sinh Le~Hong
  Nguyen, Linsheng Li, and Katsuyuki Haneda.
\newblock Hybrid {Beamforming} for {Massive} {MIMO}: {A} {Survey}.
\newblock {\em IEEE Communications Magazine}, 55(9):134--141, 2017.

\bibitem{wenxian11}
Irfan Ahmed, Hedi Khammari, Adnan Shahid, Ahmed Musa, Kwang~Soon Kim, Eli
  De~Poorter, and Ingrid Moerman.
\newblock A {Survey} on {Hybrid} {Beamforming} {Techniques} in 5g:
  {Architecture} and {System} {Model} {Perspectives}.
\newblock {\em IEEE Communications Surveys \& Tutorials}, 20(4):3060--3097,
  2018.

\bibitem{wenxian12}
Jiang Jing, Cheng Xiaoxue, and Xie Yongbin.
\newblock Energy-efficiency based downlink multi-user hybrid beamforming for
  millimeter wave massive {MIMO} system.
\newblock {\em The Journal of China Universities of Posts and
  Telecommunications}, 23(4):53--62, August 2016.

\bibitem{wenxian13}
Weiheng Ni and Xiaodai Dong.
\newblock Hybrid {Block} {Diagonalization} for {Massive} {Multiuser} {MIMO}
  {Systems}.
\newblock {\em arXiv:1504.02081 [cs, math]}, April 2015.
\newblock arXiv: 1504.02081.

\bibitem{wenxian14}
A.~Alkhateeb, G.~Leus, and R.~W. Heath.
\newblock Limited {Feedback} {Hybrid} {Precoding} for {Multi}-{User}
  {Millimeter} {Wave} {Systems}.
\newblock {\em IEEE Transactions on Wireless Communications},
  14(11):6481--6494, November 2015.

\bibitem{wenxian15}
Duy H.~N. Nguyen, Long~Bao Le, and Tho Le-Ngoc.
\newblock Hybrid {MMSE} precoding for {mmWave} multiuser {MIMO} systems.
\newblock In {\em 2016 {IEEE} {International} {Conference} on {Communications}
  ({ICC})}, pages 1--6, Kuala Lumpur, Malaysia, May 2016. IEEE.

\bibitem{wenxian3}
Jiyun Tao, Qi~Wang, Siyu Luo, and Jienan Chen.
\newblock Constrained {Deep} {Neural} {Network} based {Hybrid} {Beamforming}
  for {Millimeter} {Wave} {Massive} {MIMO} {Systems}.
\newblock {\em 2019 IEEE International Conference on Communications Workshops
  (ICC)}, pages 1--1.

\bibitem{wenxian4}
T.~J. O'Shea, K.~Karra, and T.~C. Clancy.
\newblock Learning to communicate: {Channel} auto-encoders, domain specific
  regularizers, and attention.
\newblock In {\em 2016 {IEEE} {International} {Symposium} on {Signal}
  {Processing} and {Information} {Technology} ({ISSPIT})}, pages 223--228,
  December 2016.

\bibitem{wenxian5}
X.~Gao, L.~Dai, Y.~Sun, S.~Han, and I.~Chih-Lin.
\newblock Machine learning inspired energy-efficient hybrid precoding for
  {mmWave} massive {MIMO} systems.
\newblock In {\em 2017 {IEEE} {International} {Conference} on {Communications}
  ({ICC})}, pages 1--6, May 2017.

\bibitem{wenxian6}
S.~Hur, T.~Kim, D.~J. Love, J.~V. Krogmeier, T.~A. Thomas, and A.~Ghosh.
\newblock Millimeter {Wave} {Beamforming} for {Wireless} {Backhaul} and
  {Access} in {Small} {Cell} {Networks}.
\newblock {\em IEEE Transactions on Communications}, 61(10):4391--4403, October
  2013.

\bibitem{wenxian7}
Tianqi Wang, Chao-Kai Wen, Hanqing Wang, Feifei Gao, Tao Jiang, and Shi Jin.
\newblock Deep {Learning} for {Wireless} {Physical} {Layer}: {Opportunities}
  and {Challenges}.
\newblock {\em arXiv:1710.05312 [cs, math]}, October 2017.
\newblock arXiv: 1710.05312.

\bibitem{wenxian8}
Timothy~J. O'Shea, Tugba Erpek, and T.~Charles Clancy.
\newblock Deep {Learning} {Based} {MIMO} {Communications}.
\newblock {\em arXiv:1707.07980 [cs, math]}, July 2017.
\newblock arXiv: 1707.07980.

\bibitem{R1}
G.~{Aceto}, D.~{Ciuonzo}, A.~{Montieri}, and A.~{Pescapé}.
\newblock Mobile encrypted traffic classification using deep learning:
  Experimental evaluation, lessons learned, and challenges.
\newblock {\em IEEE Transactions on Network and Service Management},
  16(2):445--458, June 2019.

\bibitem{R2}
G.~{Aceto}, D.~{Ciuonzo}, A.~{Montieri}, V.~{Persico}, and A.~{Pescapé}.
\newblock Know your big data trade-offs when classifying encrypted mobile
  traffic with deep learning.
\newblock In {\em 2019 Network Traffic Measurement and Analysis Conference
  (TMA)}, pages 121--128, June 2019.

\bibitem{wordwx2}
T.~{Lin}, J.~{Cong}, Y.~{Zhu}, J.~{Zhang}, and K.~{Ben Letaief}.
\newblock Hybrid beamforming for millimeter wave systems using the mmse
  criterion.
\newblock {\em IEEE Transactions on Communications}, 67(5):3693--3708, May
  2019.

\bibitem{wordwx3}
M.~{Joham}, W.~{Utschick}, and J.~A. {Nossek}.
\newblock Linear transmit processing in mimo communications systems.
\newblock {\em IEEE Transactions on Signal Processing}, 53(8):2700--2712, Aug
  2005.

\bibitem{wordwx4}
O.~E. {Ayach}, S.~{Rajagopal}, S.~{Abu-Surra}, Z.~{Pi}, and R.~W. {Heath}.
\newblock Spatially sparse precoding in millimeter wave mimo systems.
\newblock {\em IEEE Transactions on Wireless Communications}, 13(3):1499--1513,
  March 2014.

\bibitem{wordwx1}
F.~{Sohrabi} and W.~{Yu}.
\newblock Hybrid digital and analog beamforming design for large-scale antenna
  arrays.
\newblock {\em IEEE Journal of Selected Topics in Signal Processing},
  10(3):501--513, April 2016.

\bibitem{OMP}
M.~{Kim} and Y.~H. {Lee}.
\newblock Mse-based hybrid rf/baseband processing for millimeter-wave
  communication systems in mimo interference channels.
\newblock {\em IEEE Transactions on Vehicular Technology}, 64(6):2714--2720,
  June 2015.

\bibitem{HBF}
X.~{Yu}, J.~{Shen}, J.~{Zhang}, and K.~B. {Letaief}.
\newblock Alternating minimization algorithms for hybrid precoding in
  millimeter wave mimo systems.
\newblock {\em IEEE Journal of Selected Topics in Signal Processing},
  10(3):485--500, April 2016.

\end{thebibliography}
\end{document}